\documentclass[prl,twocolumn,superscriptaddress]{revtex4-2}

\usepackage{natbib} 
\usepackage[dvipdfmx]{graphicx} 
\usepackage{dcolumn}
\usepackage{bm}
\usepackage{color}
\usepackage{ulem}
\usepackage{lineno}

\graphicspath{{fig/}}
\usepackage{hyperref}

\begin{document}
\preprint{APS/123-QED}

\title{
Theoretical study of superconductivity in freestanding infinite-layer nickelate membranes under pressure:  mitigation of excess correlation enhances $T_c$
}

\author{Mahiru Seki}
\email{m26j3022z@edu.tottori-u.ac.jp}
\affiliation{Faculty of Engineering, Tottori University, 4-10 Koyama-cho, Tottori, Tottori 680-8552, Japan}
\author{Reo Kono}
\affiliation{Faculty of Engineering, Tottori University, 4-10 Koyama-cho, Tottori, Tottori 680-8552, Japan}
\author{Naotaka Tanaka}
\affiliation{Faculty of Engineering, Tottori University, 4-10 Koyama-cho, Tottori, Tottori 680-8552, Japan}
\author{Kensei Ushio}
\affiliation{Faculty of Engineering, Tottori University, 4-10 Koyama-cho, Tottori, Tottori 680-8552, Japan}
\author{Daiki Nakaoka}
\affiliation{Faculty of Engineering, Tottori University, 4-10 Koyama-cho, Tottori, Tottori 680-8552, Japan}
\author{Masayuki Ochi}
\affiliation{Department of Physics, Osaka University, 1-1 Machikaneyama-cho, Toyonaka, Osaka 560-0043, Japan}
\affiliation{Forefront Research Center, University of Osaka, 1-1 Machikaneyama-cho, Toyonaka, Osaka 560-0043, Japan}
\author{Kazuhiko Kuroki}
\affiliation{Department of Physics, Osaka University, 1-1 Machikaneyama-cho, Toyonaka, Osaka 560-0043, Japan}
\author{Hirofumi Sakakibara}
\email{sakakibara@tottori-u.ac.jp}
\affiliation{Faculty of Engineering, Tottori University, 4-10 Koyama-cho, Tottori, Tottori 680-8552, Japan}
\affiliation{Advanced Mechanical and Electronic System Research Center(AMES), Faculty of Engineering, Tottori University, 4-10 Koyama-cho, Tottori, Tottori 680-8552, Japan}

\date{\today}

\begin{abstract}
We theoretically investigate a freestanding membrane of infinite-layer nickelate $\mathrm{Nd}_{0.85}\mathrm{Sr}_{0.15}\mathrm{NiO}_2$ under pressure by constructing a seven-orbital effective model based on first-principles calculations.
 By performing the fluctuation exchange (FLEX) approximation, we demonstrate that the seven-orbital model explains a monotonic increase in $T_c$ reported in a recent experiment.
This enhancement of superconductivity is attributed to the mitigation of excessively strong electron correlations caused by exceptionally low valence of Ni atom.
Furthermore, we examine the dynamical stability of the crystal structure under pressure through phonon calculation.
\end{abstract}

\pacs{74.20.Mn,74.70.−b}
\maketitle

{\it Introduction.}---
High-temperature superconductivity remains one of the most significant issues in condensed matter physics. For decades, layered cuprates, in which the essential physics is believed to be captured by a single-band Hubbard model, hold the record for the highest transition temperatures ($T_c$) at ambient pressure. Furthermore, the recent discovery of superconductivity in the bilayer Ruddlesden-Popper nickelate La$_3$Ni$_2$O$_7$ under pressure \cite{MWang} has garnered intense interest, as its $T_c$ is comparable to those of the cuprates. The subsequent discoveries of La$_4$Ni$_3$O$_{10}$~\cite{sakakibara4310,NagataLa4310,2311.05453, La4Ni3O10Nature, PhysRevX.15.021005} and ambient pressure superconductivity in La$_3$Ni$_2$O$_7$ thin-film~\cite{Ko2025,QKXueThin,2504.16372,Osada2025} collectively signal the emergence of a ``nickel age'' \cite{Norman} in the study of unconventional superconductivity.

Among nickelate superconductors, the first examples of unconventional superconductors are hole-doped thin films of the infinite-layer phase, $RE_{1-x}AE_{x}$NiO$_2$ ($RE$ = La, Pr, Nd; $AE$ = Ca, Sr)~\cite{Hwang, Hwang2, Ariando, AriandoCa, OsadaSr, OsadaPr, OsadaPRM, NomuraReview}, which were studied~\cite{Pickett,Anisimov} as cuprate analog compounds $d^9$ before experimental discovery.
Interestingly, superconductivity has not yet been observed in bulk samples~\cite{HaywardNd, Hayward, Li2020,Kaneko,Kawai,Onozuka,Crespin,Ikeda}.
Since the first discovery~\cite{Hwang}, despite its relatively low $T_c$ compared to those of cuprates, infinite-layer nickelates continue to be a subject of intense investigation from a multifaceted perspective.
For example, several pathways have been identified to enhance $T_c$ through structural modifications, such as the application of compressive strain \cite{Lee2023, Ren2023} and chemical pressure \cite{Chow2025, Yang2026}. Furthermore, a very recent report on high-pressurized freestanding membranes~\cite{YanFreestanding,LeeFreestanding} of $\mathrm{Nd}_{0.85}\mathrm{Sr}_{0.15}\mathrm{NiO}_2$ have exhibited as high-$T_c$ superconductivity~\cite{YLee2026} as that of cuprates, without saturation of $T_c$. 
It is interesting that this behavior is in contrast to that of  high-$T_c$ cuprates~\cite{Hg-saturate,YBCO-saturate}.
 Consequently, it is timely to re-evaluate the electronic and superconducting properties of these systems.

While importance of additional orbital degrees of freedom such as interstitial $s$ orbitals~\cite{Nomura, Nomura2} or 5$d$ orbitals of rare-earth site~\cite{SakakibaraNi,HigashiHariki,HuZhang}, or the importance of Hund's coupling has been discussed~\cite{KotliarPRL,KotliarPRB,Xiangang,SawatzkyHS,GMZhangPRB,GMzhangFIP}, 
a number of theoretical studies argues $d$-wave pairing scenario and/or suggest effectiveness of the single orbital Hubbard model~\cite{SakakibaraNi,Kitatani, Kitatani2,Thomale,DiCataldo2024}.
$d$-wave pairing scenario is supported by recent experimental evidences~\cite{HHWen,Cheng2024,PNASdwave}.
Overall electronic structure is usually discribed as Mott-Hubbard type because of exceptionally low valence of ${\rm Ni^{1+}}$~\cite{KarpPr,HigashiHariki}, as confirmed in several experiments~\cite{CoreX,Hepting2020}.
While long range magnetic order is not observed~\cite{Hayward,HaywardNd}, a number of theories suggest 
the importance of short-range magnetic order and/or spin fluctuation similarly to cuprates~\cite{Botana,Gu2020,SakakibaraNi,Kitatani,Ryee2020,KarpCT,KarpDMFT2,LechermanMag,Leonov,Leonov-tancho,Xiangang,KotliarPRB,HuZhang,Liu2020,ZJLang,ChoiPRR}.

The origin of relatively low $T_c(\leq 20~{\rm K})$ in the infinite layer nickelates, however, is still an open question.
Present authors have proposed a scenario that excessively strong electron correlation originating from the
stronger electron repulsion suppress $T_c$ because of significant quasiparticle damping~\cite{SakakibaraNi}.
The strong repulsion is a consequence of weaker $d$-$p$ orbital hybridization than that of cuprates, which is typical in Mott-Hubbard type compounds. 
In addition, the narrower band width compared to cuprates also promotes this damping effect.
This scenario implies that reducing Ni-O bond distance should enhance $T_c$, by mitigating the correlation effects~\cite{SakakibaraNi}.
A similar view was also obtained in a dynamical vertex approximation (D$\Gamma$A) study~\cite{DiCataldo2024,KitataniDGA}.
Regarding explicitly pressurizing approach,
Wang  {\it et al.}  have reported a monotonic increase of $T_c$ in Pr$_{0.82}$Sr$_{0.12}$NiO$_2$ on SrTiO$_3$ substrates up to 12 GPa~\cite{NNWang2022}.
Notably, the previous first-principles studies have reported that electron interaction parameters remain nearly unchanged under pressure~\cite{DiCataldo2024,WernerPress}.

\begin{figure}
    \includegraphics[width=8.3cm, keepaspectratio]{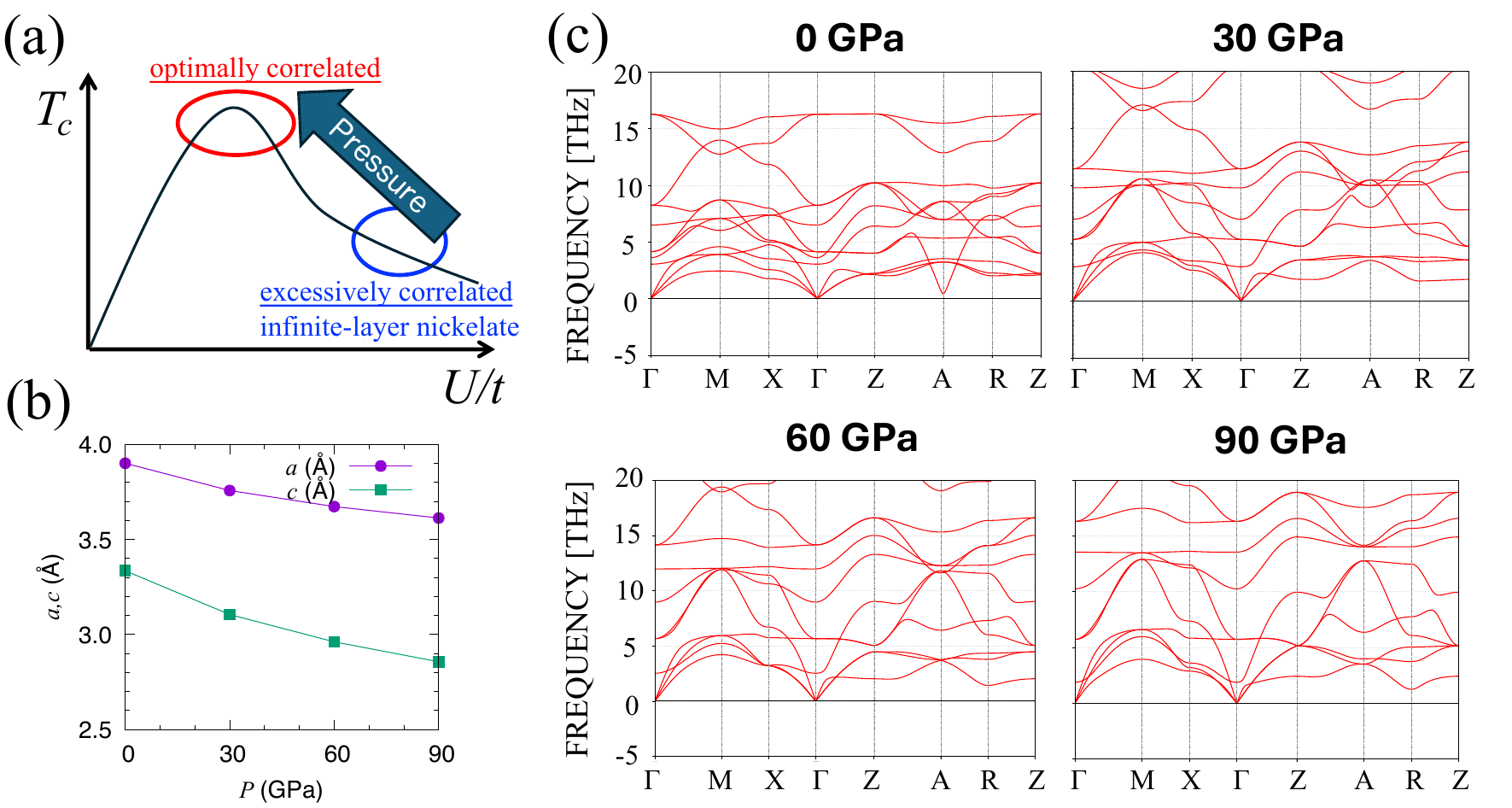}
    \caption{(a) Schematic diagram showing the gist of this study, 
    (b) lattice constant $a,c$ as functions of external pressure $P$, and (c) 
    phonon dispersion of $\rm{Nd}_{0.85}Sr_{0.15}NiO_2$ under pressure are presented.}
\label{fig1} 
\end{figure}

In the present study, we theoretically investigate freestanding membranes of $\mathrm{Nd}_{0.85}\mathrm{Sr}_{0.15}\mathrm{NiO}_2$ under pressure~\cite{YLee2026}. Specifically, we determine the energetically stable parameters of lattice constants and demonstrate that tetragonal $P4/mmm$ phase remains dynamically stable under pressures up to 90 GPa by performing phonon calculation.
Then we construct an effective model Hamiltonians via conventional Wannierization process. 
To evaluate the superconducting instability, we employ the fluctuation exchange (FLEX) approximation. 
We have concluded that the mitigation of the excessively strong correlation effects is a key to understand the enhancement of $T_c$,
as represented in Fig.~\ref{fig1}(a).

{\it Method.}---
We performed first-principles structural optimizations to determine the crystal structure using the {\footnotesize QUANTUM ESPRESSO}(QE) package~\cite{QE,QE2}.
We employed the scalar-relativistic Optimized Norm-Conserving Vanderbilt pseudopotential (ONCVPSP)~\cite{ONCVP} taken from PseudoDojo~\cite{Dojo}, parametrized using the Perdew-Burke-Ernzerhof (PBE) type generalized gradient approximation (GGA)~\cite{PBE-GGA}.
The open-core treatment was applied to Nd to represent the Nd-$f^3$ state.
A plane-wave cutoff energy of 100 Ry and a $12 \times 12 \times 12$ $k$-point mesh were used, with a Gaussian smearing width of 0.02 Ry.
Throughout this work, we assumed tetragonal $P4/mmm$ symmetry for $\rm{Nd}_{0.85}Sr_{0.15}NiO_2$. 
We adopt virtual crystal approximation (VCA) to take into account the effect of the partial substitution. 

For each case, we examined the dynamical stability by performing phonon calculations using the finite displacement method as implemented in the {\footnotesize PHONOPY} code~\cite{phonopy1,phonopy2}. 
These calculations were performed using $2 \times 2 \times 2$ supercells with a $6 \times 6 \times 6$ $k$-point mesh. 
The absence of imaginary phonon modes confirms the dynamical stability of the $P4/mmm$ structure.
In Fig. S2 of Supplemental Material, we show further checks for denser meshes~\cite{SM}.
  \par

Next, we calculated the Kohn-Sham orbitals and the electronic band dispersion based on the optimized structure. 
Based on the Kohn-Sham orbitals, we have constructed maximally localized Wannier functions \cite{Souza,Marzari}
using Wannier90 code~\cite{wannier90_pizzi,wannier90_mostofi},
by which we also obtain the hopping parameters among the Wannier functions. 
We have constructed a seven-orbital model consisting of the all Ni-$3d$ orbitals and Nd-$5d_{3z^2-r^2},d_{xy}$ orbitals, as in Ref.~\cite{SakakibaraNi,modelcomment}.
We adopt constrained random phase approximation (cRPA~\cite{cRPA}) to the model, for determining the onsite Coulomb $(U,U')$ and exchange interaction $(J,J')$ parameters of the model using the {\footnotesize RESPACK} codes~\cite{respack0,RESPACK,RESPACK1,RESPACK2,RESPACK3,RESPACK4,RESPACK5}.
During these process, Wannier functions obtained by Wannier90 has been formatted to be that of {\footnotesize RESPACK} by utilizing {\footnotesize WAN2RESPACK} code~\cite{wan2respack}.
We took a plane-wave cutoff energy of 100 Ry and a $10 \times 10 \times 10$ $k$-point mesh for Wannerization and cRPA.
The energy cutoff for the dielectric function was set to be 25 Ry.
The complete set of these parameters are summarized in Table S1 of Supplemental Material~\cite{SM}.
\par

We explored the possibility of superconductivity for the obtained low-energy seven-orbital model within FLEX approximation~\cite{Bickers,Bickers1991}. 
We calculated the self-energy induced by the spin-fluctuation formulated as shown in the literatures~\cite{Lichtenstein,mFLEX1,mFLEX2} in a self-consistent calculation.
The explicit formulae of the irreducible, spin, and charge susceptibilities 
describing the fluctuations are shown in Eqs.(2)-(4) of Ref.~\cite{Sakakibara2}.
The real part of the self-energy at the lowest Matsubara frequency was subtracted in the same manner with Ref.~\cite{Ikeda_omega0} to maintain the band structure around the Fermi level obtained by first-principles calculation.

The obtained Green's function and the pairing interaction, mediated mainly by spin fluctuations, were plugged into the linearized Eliashberg equation.
Since the eigenvalue $\lambda$ of the linearized Eliashberg equation reaches unity at $T=T_c$, we adopted it as a measure of superconductivity at a fixed temperature, $T=0.01$ eV. 
For convenience, we will call the eigenfunction (with the largest eigenvalue) of the linearized Eliashberg equation at the lowest Matsubara frequency $i\omega$(=$i\pi k_{\rm B}T$) the ``superconducting gap function''. 
In the present study, the eigenfunctions of largest $\lambda$ are  $d_{x^2-y^2}$-wave (denoted as $d$-wave in the followings for simplicity) for all cases.
We took an 8$\times$8$\times$8 $k$-point mesh and 8192 Matsubara frequencies for the FLEX calculation. 


{\it Results}.---
In Fig.~\ref{fig1}(b), we show the variation of lattice parameters $a,c$ induced by external pressure.
There are clear decrease in both parameters.
The constant $c$ decreases rather faster, this may be because of more spatiality along $c$ axis in the unit cell. 
In Fig.~\ref{fig1}(c), we also plot the phonon dispersion for the cases at $P=0,30,60,90$ GPa, where the absence of imaginary mode certifies the dynamical stability of $P4/mmm$ structure.
This result indeed suggests that our following analysis about superconductivity assuming $P4/mmm$ structure is plausible.

\begin{figure}
    \includegraphics[width=8.5cm, keepaspectratio]{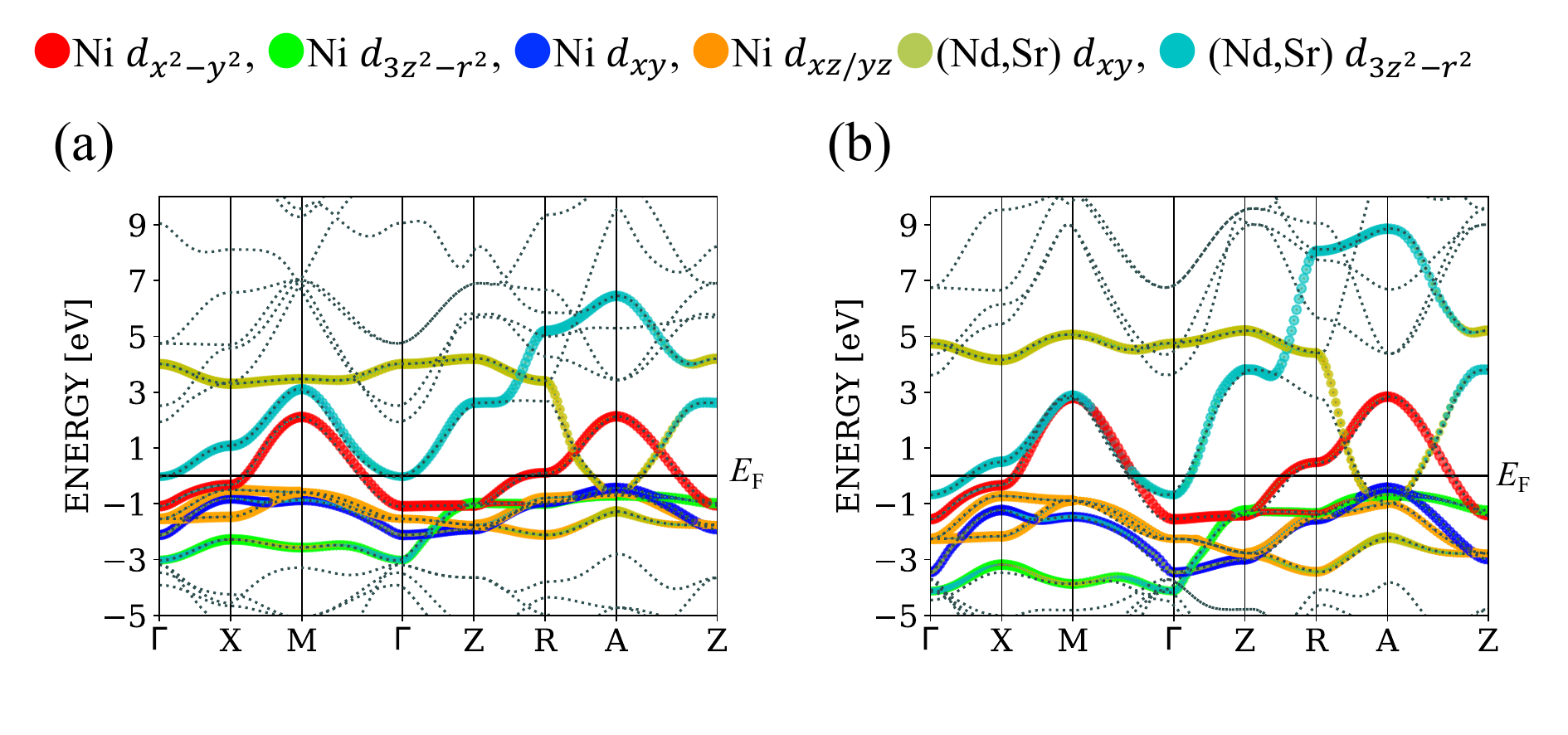}
    \caption{Band structure with respect to the Fermi level ($E_F=0$) of $\rm{Nd}_{0.85}Sr_{0.15}NiO_2$ under pressure for the cases of (a) 0 GPa and (b) 90 GPa.
    Here, the weight of Wannier orbitals are represented by the thickness of color coded lines.
    }
\label{fig2} 
\end{figure}

In Fig.~\ref{fig2}, we plot the band structure, where the weight of Wannier orbitals are superposed.
Here the ambient pressure case is basically the same with our previous calculation for the 20\% doped case ~\cite{structcomment,SakakibaraNi}.
We find increase of the overall band width 
 as pressure increases, consistent with previous papers~\cite{DiCataldo2024,CanoGW2}.
 This change is basically interpreted as a consequence of reduction of parameters $a,c$.
Especially, three dimensionality is evidently increased by shorten $c$.
This change is found in the comparison between $\Gamma$-X-M and Z-R-A $k$-paths. 
The crucial changes for superconductivity are these two:
the Fermi pockets consisting of $5d$ orbitals become large and the van Hove singularity of the $3d_{x^2-y^2}$ orbital shift upward, where the latter changes the property of nesting.

\begin{figure}
    \includegraphics[width=8.4cm, keepaspectratio]{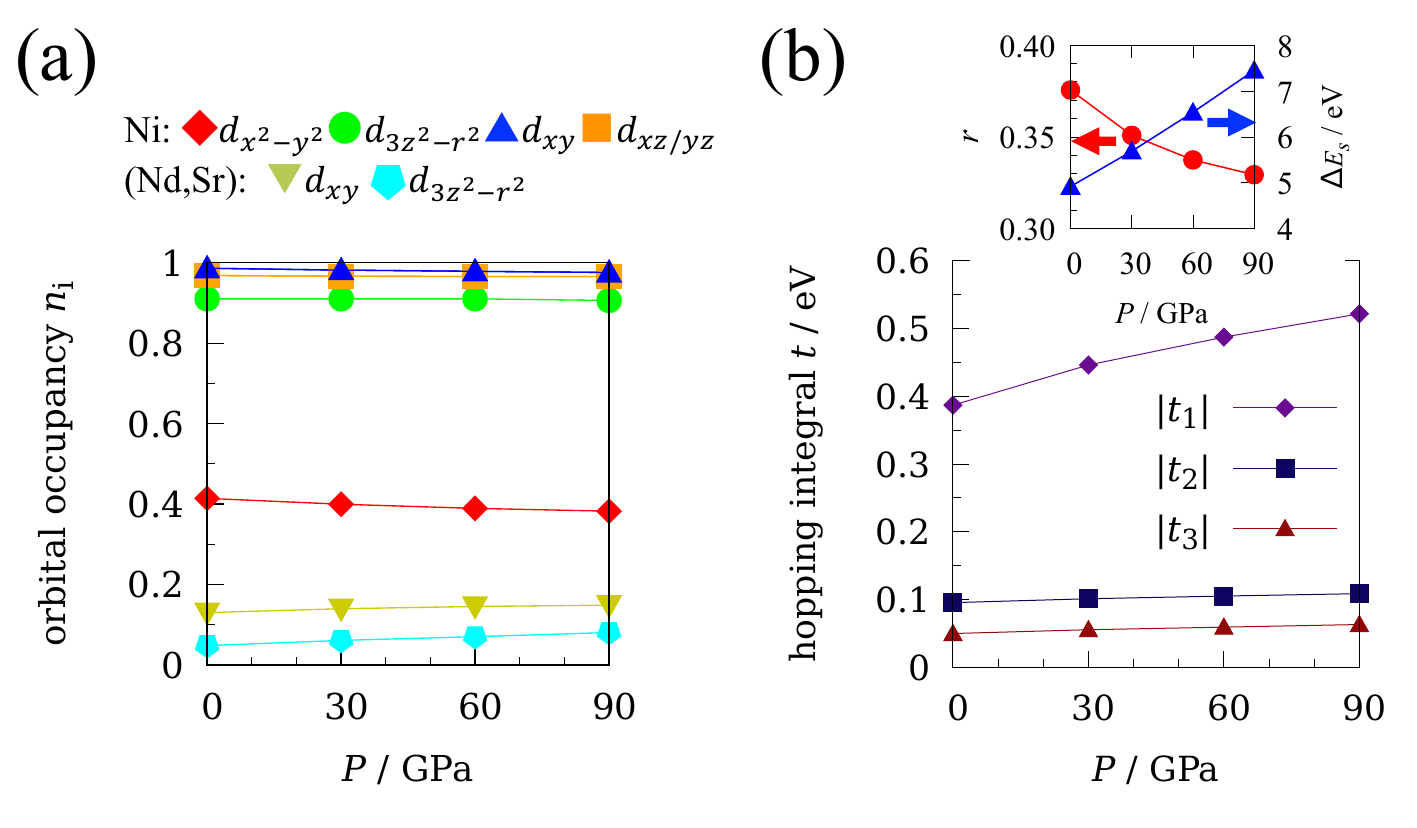}
    \caption{
    (a) Orbital occupancy of Wannier orbitals and (b) hopping integral $t$ among the near neighbor Wannier $d_{x^2-y^2}$ orbital as functions of external pressure $P$, where the destinations in the hopping are indicated.  In the inset, the roundness parameter $r=(|t_2|+|t_3|)/|t_1|$ of the Fermi surface (red circle) and $\Delta E_s=E_{4s}-E_{3dx^2-y^2}$ (blue triangle) are plotted. 
    }
\label{fig3} 
\end{figure}

In Fig.~\ref{fig3}(a), we plot occupancies of Wannier orbitals,
 where the total number of electrons per unit cell $n$ is fixed at 4.425 (corresponding to 15\% substitution of Sr at Nd site).
The orbital occupancy $n_{x^2-y^2}$ of the $d_{x^2-y^2}$ Wannier orbital
decreases as pressure increases, as discussed in the previous study~\cite{DiCataldo2024}(Tab. S1).
The total number of electrons is compensated by the increase of Fermi pockets in size, which consist of mixed states between the $3d_{yz/xz}$ and $5d_{xy}$($3d_{3z^2-r^2}$ and $5d_{3z^2-r^2}$) orbitals around $\Gamma$(A) points in the Brillouin zone (namely, the effect of self-doping~\cite{Pickett,SakakibaraNi,CanoGW1}).
The decrease of $n_{x^2-y^2}$ basically weakens the spin-fluctuation and tendency toward the $d$-wave superconductivity by shifting the $d_{x^2-y^2}$-orbital band from the optimally doped region toward overdoped region.

In Fig.~\ref{fig3}(b), we plot the absolute values of the $i$-th nearest neighbor hopping integrals $t_i$ ($i=1,2,3$) as a function of applied pressure.
Interestingly, while $t_1$ significantly increases, $t_2,t_3$ barely increase.
This can be explained by the effect of $4s$ orbital~\cite{Pavarini,Andersen} of Ni atom:
it enhances $t_2,t_3$ through a second order perturbation process, as discribed in Fig. 5 in Ref.~\cite{Sakakibara2}.
Namely, as the $4s$ orbital is repulsed by the surrounding oxygens  more strongly than the $3d_{x^2-y^2}$ orbitals, pressure weakens the hybridization between $4s$ and $3d_{x^2-y^2}$ orbitals, which leads to larger $t_2,t_3$.
This change gives rise to a stronger spin fluctuation among the $3d_{x^2-y^2}$ orbitals by suppressing the roundness of the Fermi surface.
We quantify this tendency defining roundness parameter $r=(|t_2|+|t_3|)/|t_1|$~\cite{Sakakibara3}  as a relative strength among $t_i$.
We see a systematic suppression of $r$ in the inset of Fig.~\ref{fig3} (b),
which tendency is similar to the case of HgBa$_2$CuO$_4$ under pressure~\cite{SakakibaraP}.


\begin{figure}
    \includegraphics[width=8.5cm, keepaspectratio]{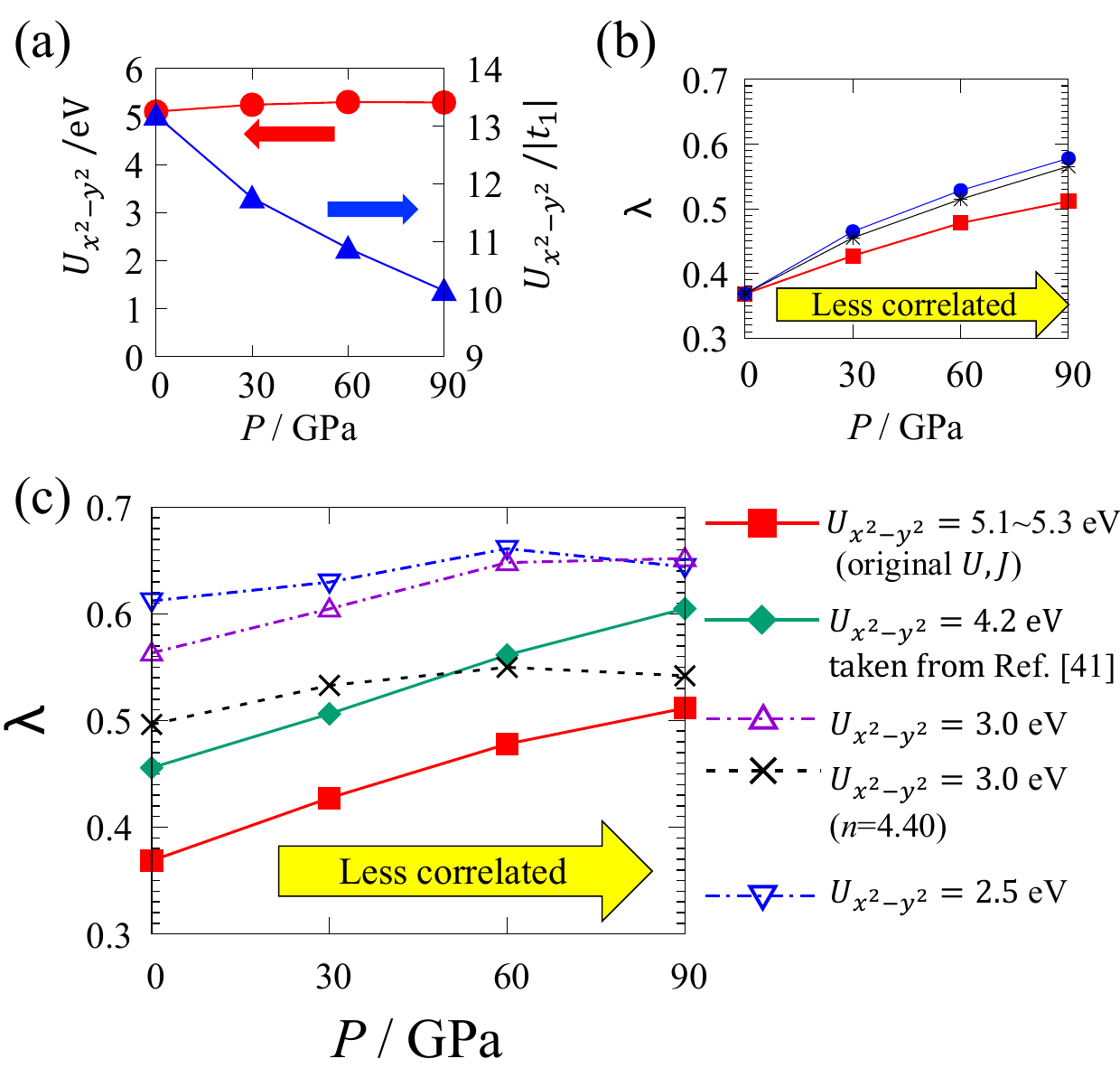}
    \caption{Onsite intraorbital interaction $U_{x^2-y^2}$ (red circle, panel (a)), the ratio $U_{x^2-y^2}/|t_1|$ (blue triangle), and
    $\lambda$ of $d$-wave superconductivity (panel (b) and (c)) are plotted as a function of pressure $P$. 
    In panel (b), the original case (red squares) and the hypothetical cases equating only the parameter set $(t_1,t_2,t_3)$ (blue circle)
    and only $(t_1,t_2,t_3,U,U',J,J')$ (black asterisks) with those of the pressurized model are shown. 
    In panel (c), green diamonds indicate the results obtained by using the value of Ref.~\cite{SakakibaraNi}.
    The open upward (downward) triangles indicate the $U_{x^2-y^2}=3.0$ and 2.5 eV cases (see text), respectively.
     The cross symbols indicate those of $U_{x^2-y^2}=3.0$ eV at $n=4.4$, while the other cases are calculated at $n=4.425$. 
     }
\label{fig4}
\end{figure}

Now we come to superconductivity. 
In Fig.~\ref{fig4}(a), we plot the value of intraorbital interaction $U_{x^2-y^2}$ of the $d_{x^2-y^2}$ orbital and the ratio $U_{x^2-y^2}/|t_1|$ as a function of pressure.
Since $U_{x^2-y^2}$ remains almost unchanged while $|t_1|$ clearly increases, the ratio $U_{x^2-y^2}/|t_1|$ decreases significantly.
In Fig.~\ref{fig4} (b), the eigenvalues $\lambda$($d$-wave) of the Eliashberg equation is plotted as a function of pressure. 
The monotonically increasing trend of $\lambda$ is a consequence of the decrease of $U_{x^2-y^2}/|t_1|$, and it is fairly consistent with that of experimental $T_c$~\cite{YLee2026}.
To directly verify the origin of increasing $\lambda$, we re-performed the FLEX calculation for a hypothetical model in which only $t_1, t_2,$ and $t_3$ were set equal to those of the pressurized models (Fig.~\ref{fig4}(b)), confirming a significant increase in $\lambda$. We also examined the case where $(U, U', J, J')$ were equated to the pressurized values, which revealed that the variation in interaction parameters is less effective. The main difference between the original and hypothetical models is likely attributable to the self-doping effect, as discussed in Fig.~\ref{fig3}.
In Fig.~\ref{fig4}(c), we have also plotted $\lambda$ calculated by using the parameter set of $(U,U',J,J')$ presented in Ref.~\cite{SakakibaraNi}, where $U_{x^2-y^2}\simeq 4.2$ eV, smaller in nearly 1 eV than that of the present calculation ($\simeq$ 5.1 eV at ambient pressure, Fig.~\ref{fig4}(a)). This comparison also demonstrates that the smaller $U_{x^2-y^2}/|t_1|$ is favorable for superconductivity,
namely, the mitigation of electron correlation enhances $T_c$ at least within this $U_{x^2-y^2}/|t_1|$ regime.

We have also studied cases with smaller values of $U_{x^2-y^2}/|t_1|$.
Namely, we also plot in Fig.~\ref{fig4}(c) $\lambda$ for the cases of $U_{x^2-y^2}=3.0$ eV and $2.5$ eV where the ratio among values of $(U,U',J,J')$ are kept to be the same with the original values. The rapid saturation of $\lambda$ against $P$ and even the dome-shape is in disagreement with the experiment~\cite{YLee2026}. This is because by adopting such small values of $U_{x^2-y^2}$, $U_{x^2-y^2}/|t_1|$ reaches its optimum value before the pressure reaches 90 GPa.  Hence, only the realistic values of $U_{x^2-y^2}=4-5$ eV give results consistent with the experiment~\cite{YLee2026}.
In Fig.~\ref{fig4} (b), we also compare the cases of two different band filling ($n=4.425$ and $n=4.4$) to find that larger amount of hole-doping strengthens non-monotonic trends.


{\it Discussion.}---
Regarding the first-principles estimation of $U_{x^2-y^2}$, the present value of $U\sim 5.1$ eV at ambient pressure is close to that of Ref.~\cite{Nomura}(Tab. V).
In Table S2 of Supplemental Material~\cite{SM}, we report the values calculated by Vienna Ab initio Simulation Package (VASP)~\cite{vasp1,vasp2,vasp3,vasp4}, a basically plane wave method as same with QE, is consistent with our result.
On the contrary, the data of Linear Muffin-Tin Orbital Method (LMTO) based codes in Refs.~\cite{SakakibaraNi,DiCataldo2024}
are close to each other, implying that the ambiguity of $U_{x^2-y^2}$ is given by the basis set used in the first-principles calculation.
This difference might also originate from the choice of lanthanoide element, where Nd is used in this study while La was used in the previous study~\cite{SakakibaraNi}. A comparison between them is shown in the supplemental material of the previous paper~\cite{SakakibaraNi}.

\begin{figure}
    \includegraphics[width=7cm, keepaspectratio]{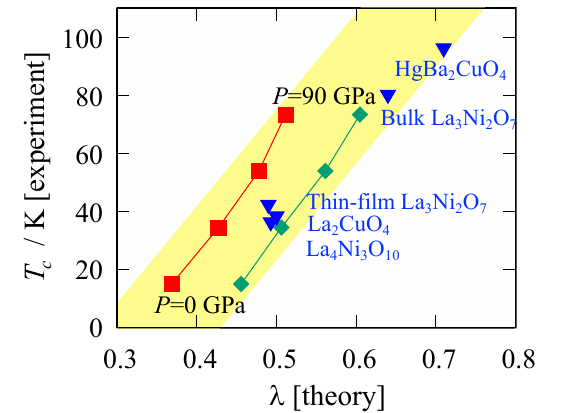}
    \caption{$T_c$'s are plotted against the eigenvalues of Eliashberg equation $\lambda$ at a fixed temperature, $T=0.01$ eV.	
    Red squares and green diamonds indicate the same $\lambda$ with that of Fig.~\ref{fig4}(b).
    Blue downward triangles indicate the data of cuprates and Ruddlesden-Popper type nickelates, calculated in the same method as described in the Appendix of Ref.~\cite{Ushio}.	Yellow hatched region is a guide to the eyes.  
      }
\label{fig5}
\end{figure}

In Fig.~\ref{fig5}, we plot the correlation between $\lambda$ and experimentally observed $T_c$, where the detailed information of the calculation is presented in Supplemental Material~\cite{SM}.
Although there seems to be better consistency among materials in the case of LMTO(by ecalj code~\cite{ecalj}) than that of the plane wave method, the overall trends are basically consistent in both methods.
This suggests that our present study gives a good description for the enhancement of $T_c$ in freestanding membrane of nickelate~\cite{YLee2026}.

{\it Summary.}---
To summarize, we have performed first-principles structural optimization and phonon calculations to determine the energetically and dynamically stable crystal structures of a free-standing $\mathrm{Nd}_{0.85}\mathrm{Sr}_{0.15}\mathrm{NiO}_2$ membrane under pressure.
We evaluated the electron interaction parameters $(U, U', J, J')$ for the seven-orbital model constructed based on the obtained structure, and found that $U_{x^2-y^2}/|t_1|$ significantly decreases in the high-pressure region.
By applying FLEX approximation to this model, we demonstrated a monotonic increase in the eigenvalue $\lambda$, which is consistent with the experimental enhancement of $T_c$~\cite{YLee2026}.
This agreement suggests that mitigating excessively strong electron correlations enhances $T_c$, and that a model with a sufficiently large $U_{x^2-y^2}/|t_1|$ is more plausible for infinite-layer nickelates than one with smaller values.

\begin{acknowledgments}
The computing resource is supported by 
the supercomputer system (system-B) in the Institute for Solid State Physics, the University of Tokyo, 
and the supercomputer of Academic Center for Computing and Media Studies (ACCMS), Kyoto University.
This work was supported by JST K Program Grant No. JPMJKP25Z3, and Grants-in-Aid for Scientific Research from
JSPS, KAKENHI Grants No. JP24K01333, JP25H01252, JP25K08459 (H.S.), JP25K00959, JP26K08179 (K.K).
M.O. and H.S. were supported by JST FOREST Program, Grant No. JPMJFR212P(M.O.) and JPMJFR246T(H.S.).
\end{acknowledgments}

\bibliography{srnio2}

\begin{thebibliography}{116}
\bibitem [{mod()}]{modelcomment}%
  \BibitemOpen
  \href@noop {} {}\bibinfo {note} {This simplification of incorporating the O
  2$p$ orbitals in the ``$d$-only'' model is often adopted in studies of
  cuprate by introducing the concept of Zhang-Rice singlet~\cite{ZR}. As
  discussed in Ref.~\cite{SakakibaraNi}, $d$-only model becomes even more
  natural due to the weaker $d$-$p$ hybridization than in the
  cuprates.}\BibitemShut {Stop}%
\bibitem [{str()}]{structcomment}%
  \BibitemOpen
  \href@noop {} {}\bibinfo {note} {In Ref.~\cite{SakakibaraNi}, we have
  calculated (La,Ba)NiO$_2$ adopting experimental lattice parameters of
  NdNiO$_2$~\cite{HaywardNd} to mimic the band structure of (Nd,Sr)NiO$_2$ for
  technical reasons.}\BibitemShut {Stop}%
\end{thebibliography}

\end{document}